\documentclass[
 reprint,
 amsmath,amssymb,
 pra,
 superscriptaddress,
]{revtex4-1}
\usepackage{graphicx}
\usepackage{dcolumn}
\usepackage{bm}

\begin{document}

\title{Weyl solitons in three-dimensional optical lattices}

\author{Ce Shang}
\email{shangce1989@sjtu.edu.cn}
\affiliation{The State Key Laboratory of Advanced Optical Communication Systems and Networks, Shanghai Jiao Tong University, Shanghai 200240, China}
\affiliation{Key Laboratory for Laser Plasma (Ministry of Education), Collaborative Innovation Center of IFSA, School of Physics and Astronomy, Shanghai Jiao Tong University, Shanghai 200240, China }
\author{Yuanlin Zheng}
\affiliation{The State Key Laboratory of Advanced Optical Communication Systems and Networks, Shanghai Jiao Tong University, Shanghai 200240, China}
\affiliation{Key Laboratory for Laser Plasma (Ministry of Education), Collaborative Innovation Center of IFSA, School of Physics and Astronomy, Shanghai Jiao Tong University, Shanghai 200240, China }
\author{Boris A. Malomed}
\affiliation{Department of Physical Electronics, School of Electrical Engineering, Faculty of Engineering, Tel Aviv University, Tel Aviv 69978, Israel }

\begin{abstract}
Weyl fermions are massless chiral quasiparticles existing in materials known as Weyl semimetals. Topological surface states, associated with the unusual electronic structure in the Weyl semimetals, have been recently demonstrated in linear systems. Ultracold atomic gases, featuring laser-assisted tunneling in three-dimensional optical lattices, can be used for the emulation of Weyl semimetals, including nonlinear effects induced by the collisional nonlinearity of atomic Bose-Einstein condensates. We demonstrate that this setting gives rise to topological states in the form of Weyl solitons at the surface 
of the underlying optical lattice. These nonlinear modes, being exceptionally robust, bifurcate from linear states for a given quasi-momentum. The Weyl solitons may be used to design an efficient control scheme for  topologically-protected unidirectional propagation of excitations in light-matter-interaction physics. After the recently introduced Majorana and Dirac solitons, the Weyl solitons proposed in this work constitute the third (and the last) member in this family of topological solitons.
\end{abstract}

\maketitle

\section{Introduction}

Three species of fermions, of the Dirac \cite{ref1}, Majorana \cite%
{ref2,ref3} and Weyl types \cite{ref4,ref5,ref6}, are cornerstones of the
relativistic quantum-field theory. On the other hand, Weyl semimetals, which
enable the realization of the Weyl fermions, is opening up a new chapter of
condensed-matter physics. Weyl points, the signature of the respective
topological charge, are produced by the Hamiltonian $H={v_{x}}{k_{x}}{\sigma
_{x}}+{v_{y}}{k_{y}}{\sigma _{y}}+{v_{z}}{k_{z}}{\sigma _{z}}$, where ${v_{j}%
}$, ${k_{j}}$ and ${\sigma _{j}}$, with $j=x,y,z$, are group velocities,
momentum components, and Pauli matrices, respectively. With linear
dispersion in all the three dimensions in its vicinity \cite%
{ref7,ref8,ref9,ref10,gao2016photonic}, these nodal points in the momentum
space are realized as magnetic monopoles in the $\mathbf{k}$-space. The
topological invariant of the Weyl semimetal may be determined by the sign
chirality, defined as $\chi=\mathrm{sign}\left( {{v_{x}}{v_{y}}{v_{z}}}\right) $%
, or the integral of the Berry curvature on a closed manifold enclosing the
Weyl point. The gapless topological states built of bulk low-energy
electrons also feature the existence of the Weyl semimetal. Aside from
solid-state electronic materials, the rapid development of the technique
based on synthetic magnetic fields in ultracold atomic gases \cite%
{goldman2013direct,celi2014synthetic} and possibilities to precisely control
properties of Bose-Einstein condensates (BECs) \cite{nalitov2015polariton}
offer an efficient platform for investigating topological phenomena and
novel states of matter. In particular, lattice models make it possible to
handle pseudospin components of the wave functions by modifying the lattice
geometry, which also may result in the emergence of Weyl points \cite%
{ref11,ref12,PhysRevX.7.041026}. The unusual structure of the wave functions near these points
gives rise to a plenty of notable topological properties and stimulates
ongoing research in various fields of physics.

The topological phenomena and relativistic particles mentioned above are
generally produced by linear systems. Nonlinearity also essentially affects
a variety of phenomena in physics, such as coherent control of excitations
\cite{ref13}, bistability \cite{gibbs2012optical}, soliton formation
\cite{ref15,lederer2008discrete,dauxois2006physics,yang2010nonlinear}%
, harmonic generation and frequency conversion \cite{ref17}, and many
others. Effects of nonlinearity on edge states, including prediction of
solitons, were recently studied in topological insulators \cite%
{ref18,ref19,kartashov2016modulational}. However, the impact of
nonlinearities on Weyl semimetals, associated with either nonlinear effects
in surface-state propagation or inter-particle interactions was not explored
yet. In the present work, we address this issue, considering competing
repulsive and attractive interactions \cite%
{quiroga1997stable,kevrekidis2005vector,carretero2008nonlinear,abdullaev2008localized,
reyna2014two,PhysRevLett.118.230403} between atoms in synthetic magnetic fields \cite%
{lin2009synthetic,dalibard2011colloquium}, which is similar to the interplay
of self-defocusing and focusing nonlinearities in optics. The objective of
our analysis is to build soliton-like surface states, in the full
three-dimensional (3D) form, in optical Weyl lattices emulating the Weyl
metals. These nonlinear lattices can be utilized to study interplay
between the nonlinearity and topologically protected surface states. We find
that soliton modes bifurcate from linear periodic surface states. We also
find that these Weyl solitons may travel along the surface without notable
deformations, featuring extremely low radiation loss. The Weyl solitons,
after the very recently reported Dirac solitons \cite{ref25} and Majorana solitons\cite{zou2016traveling},
are the third and also the last member in the family of topological solitons, finally hosting a reunion in nonlinear physics.
\begin{figure*}[htp]
\centering
{\includegraphics[width=0.8\linewidth]{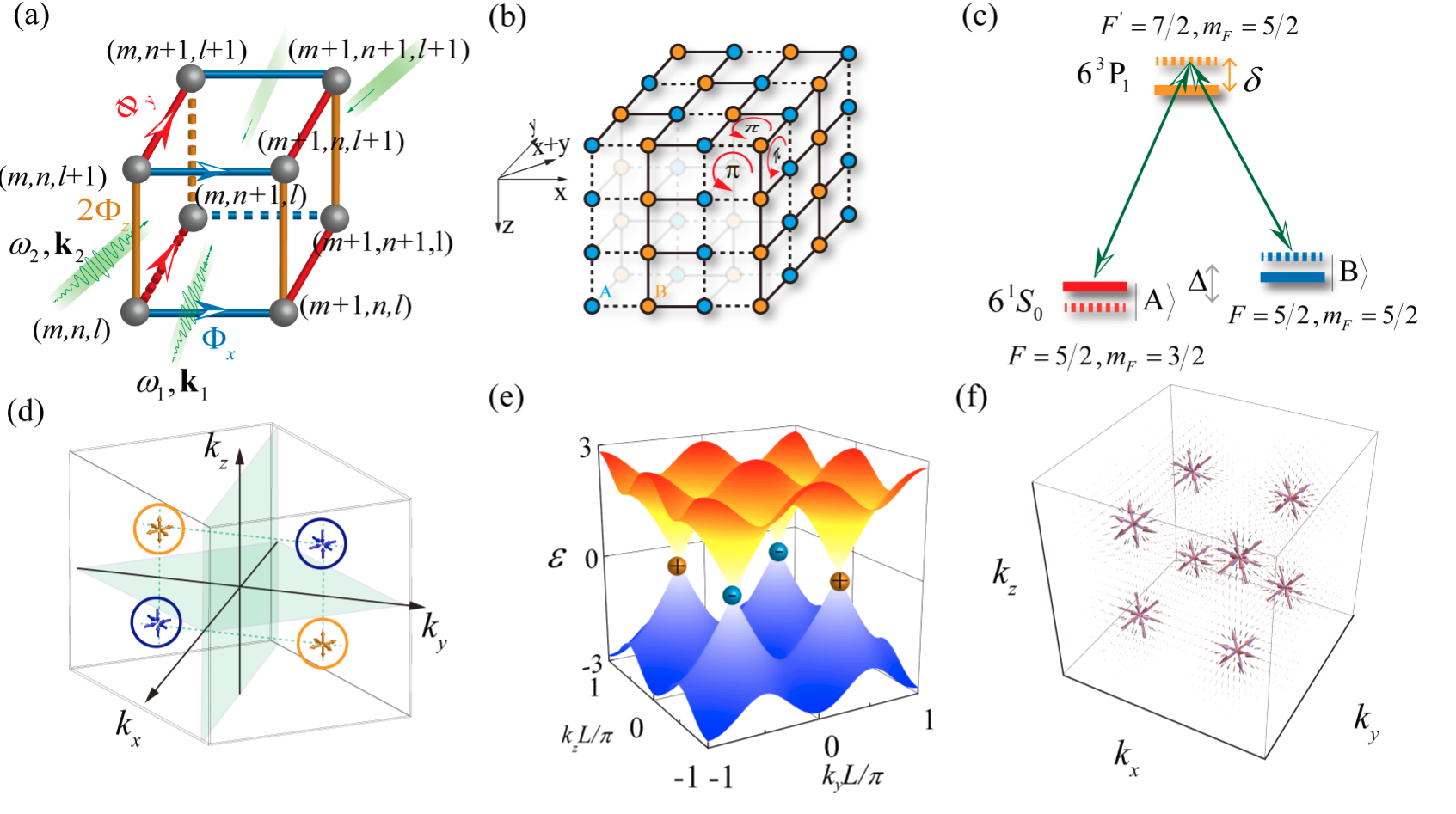}}
\caption{ (a) Optical-lattice potentials formed by superimposing two
 standing waves, and the configuration of gauge fields defined on
bonds of the cubic lattice with the corresponding coordinates. (b) A sketch
of the 3D cubic lattice with phase-driven hopping along $x$ and $z$
directions, which possesses Weyl points in the momentum space. Amplitudes of
tunneling along the lattice sites alternately carry a complex phase shown as
dashed (solid) lines, illustrating hopping with an acquired phase $\protect%
\pi $ ($0$). The unit cell can be constructed, using two pseudospin sites
marked by orange and blue colors. (c) The Raman coupling scheme for engineering
the required tunneling along each axis. The detuning $\Delta$ matches
the frequency offset of the corresponding Raman beams. (d) The positions of the Weyl points in
the Brillouin zone of the 3D Hamiltonian in the reciprocal lattice and their
chiralities are indicated by blue and orange arrow sets. (e) The band
structure with signs of the Weyl points in the $\left( {k_{y}},{k_{z}}%
\right) $ plane, at ${k_{x}}=0$. (f) Distribution of the Berry curvature of Weyl nodes
 in momentum space with the opposite chiralities.}
\label{fig:1}
\end{figure*}
\section{The linear model}

\label{sec:2}

Gauge fields play an essential part in various areas of modern physics \cite%
{ref20}. In particular, the Harper Hamiltonian in the tight-binding limit,
based on gauge fields, describes the dynamics of particles in a magnetic
field with a background lattice. This Hamiltonian has been already shown to
exhibit a fascinating fractal band structure which is called the
\textquotedblleft Hofstadter butterfly" \cite{ref21}. The particular Harper
Hamiltonian was recently proposed for the realization of the Weyl states
\cite{ref22} in ultracold atoms.

An optical Weyl lattice may be induced by the electric dipole interaction
between atoms and the electric field of an optical standing wave. As shown
in Fig. \ref{fig:1}(a), two running-wave Raman laser beams with frequencies
and wave vectors $\left[ {{\omega _{\alpha}},{\mathbf{k}_{\alpha}}}\right] $, ${\alpha=1,2}$, induce
a spatially dependent complex tunneling by the position-dependent
modulation, and generate strong synthetic magnetic fields. Due to this
modulation, the tunneling matrix element consequently picks up a
Peierls phase, which originates from the propagator of an electron in the
magnetic field. With the precise control of the laser field, one can
efficiently engineer the Peierls phase and redefine atom hopping by adding
an Aharonov-Bohm-like phase, $\Phi $, which is equal to the sum of the
Peierls phases, accumulated in the course of tunneling around the plaquette
\cite{RevModPhys.89.011004}. The plaquette, regarded as a gauge invariant,
is defined modulo the dimensionless magnetic flux quantum of $2\pi$ piercing
a lattice. In Fig. \ref{fig:1}(b), we introduce a uniform fully tunable
effective flux and the lattice built of two sublattices ($A$ and $B$, which
build a unit cell), which give rise to pseudospins with opposite magnetic
moments. Our coupling scheme directly creates a non-Abelian $SU(2)$ gauge
field that results in opposite magnetic fields for quasiparticles. Thus, a
3D cubic lattice can be constructed with laser-assisted tunneling along $x
$ and $z$ directions. For resonant tunneling \cite{jaksch2003creation,miyake2013realizing},
the time averaging over the rapidly oscillating terms
yields an effective 3D Hamiltonian:
\begin{equation}
\begin{split}
H& =-\sum\limits_{\mathbf{r},\sigma,\tau}({J_{x}}{e^{-i{\Phi_{\mathbf{r}+%
\hat x,\mathbf{r}}}}}a_{\mathbf{r} +\hat x,\sigma}^{\dag} {a_{\mathbf{r},\tau}}%
+{J_{y}}a_{\mathbf{r}+\hat y,\sigma}^{\dag} {a_{\mathbf{r},\tau}} \\
& +{J_{z}}{e^{-i{\Phi _{\mathbf{r}+\hat z,\mathbf{r}}}}}a_{\mathbf{r} +%
 \hat z,\sigma}^{\dag }{a_{\mathbf{r},\tau }}+\mathrm{H}\mathrm{.c}\mathrm{.}).
\end{split}
\label{eq:refname1}
\end{equation}%
Here, ${J_{x,y,z}}$ are tunneling amplitudes, $\sigma ,\tau \mathrm{\ =}A,B $ label %
 the two pseudospin components, $a_{\mathbf{r},\sigma,\tau}^{\dag }$ and $%
a_{\mathbf{r},\sigma,\tau} $ are the creation and annihilation operators on site $\mathbf{r}=(m,n,l)$,
and ${\Phi _{\mathbf{r}}}=\delta \mathbf{k}\cdot {R_{m,n,l}}=m{\Phi _{x}}+n{\Phi
_{y}}+l{\Phi _{z}}$ denote the nontrivial hopping phases, associated with
positions ${R_{m,n.l}}\mathrm{=}mL\hat{x}+nL\hat{y}+lL\hat{z}$ and momentum
difference $\delta {\mathbf{k}}\mathrm{=}{\mathbf{k}_{1}}-{\mathbf{k}_{2}}$,
where $m,n,l$ are integers, and $\hat{x}\,,\hat{y}\,,\hat{z}$ are unit
vectors along the $x,y,z$ direction, respectively. The lattice spacing is
$L$. We select the appropriate directions such
that $\left( {{\Phi _{x}},{\Phi _{y}},{\Phi _{z}}}\right) =\pi \left( {1,1,2}%
\right) $.

The experiment can be performed in quantum degenerate Fermi gases of $%
\mathrm{{}^{173}Yb}$ in the presence of a uniform magnetic field $\mathcal{B}
$ \cite{qi2011bound,cappellini2014direct}. We
consider the $\mathrm{6}{}^{1}{S_{0}}\rightarrow \mathrm{6}{}^{3}{\mathrm{P}%
_{1}}$ transition with one excited state $\left\vert {{F^{^{\prime }}}={7%
\mathord{\left/{\vphantom {7 2}} \right.\kern-\nulldelimiterspace}2},{m_{F}}=%
{5\mathord{\left/{\vphantom {5 2}} \right.\kern-\nulldelimiterspace}2}}%
\right\rangle $ and two ground states $\left\vert {F={5\mathord{\left/{%
\vphantom {5 2}} \right.\kern-\nulldelimiterspace}2},{m_{F}}={3%
\mathord{\left/{\vphantom {3 2}} \right.\kern-\nulldelimiterspace}2}}%
\right\rangle $, $\left\vert {F={5%
\mathord{\left/{\vphantom {5 2}}
\right.\kern-\nulldelimiterspace}2},{m_{F}}={5%
\mathord{\left/{\vphantom {5
2}}\right.\kern-\nulldelimiterspace}2}}\right\rangle $. ,
 We set the frequency difference ${\omega
_{1}}-{\omega _{2}}\approx \Delta $ $(\Delta={g_{F}}{\mu _{\mathcal{B}}}\mathcal{B})$,
so that the dressed ground states are
nearly degenerate, ${g_{F}}$ is the hyperfine Land\'{e} factor, ${\mu _{\mathcal{B}}}$ is the Bohr magneton, and ${m_{F}}$ is the projection of the atomic angular
momentum along the magnetic field. The two-photon detuning is $\delta$. In
the presence of large detuning $\Delta \gg \Omega ,\delta $, we can
adiabatically eliminate the excited state and consider the atomic motions in
the ground-state manifold, to derive the hopping term $J={{{\Omega ^{2}}}%
\mathord{\left/ {\vphantom {{{\Omega ^2}} \delta }} \right.
 \kern-\nulldelimiterspace}\delta }$, where $\Omega $ is the strength of the
Raman coupling between the ground and excited states. 

It is relevant to mention here that the proposed scheme may be actually realized
more straightforwardly for bosonic systems, such as chiral vortices in an interacting bosonic quantum fluid \cite{vortex}. Relevant details of the bosonic setting will be considered in detail elsewhere.

In the quasi-momentum representation, the present setting amounts to the 3D
Harper Hamiltonian for two sublattices: $H\left( \mathbf{k}\right) =-2\left[
{{J_{y}}\cos \left( {{k_{y}}L}\right) {\sigma _{x}}+{J_{x}}\sin \left( {{%
k_{x}}L}\right) {\sigma _{y}}-{J_{z}}\cos \left( {{k_{z}}L}\right) {\sigma
_{z}}}\right] $. In the first Brillouin zone (BZ) in Fig. \ref{fig:1}(d) the
two energy bands $\varepsilon \left( k\right) =\pm 2\sqrt{J_{x}^{2}{{\sin }%
^{2}}\left( {{k_{x}}L}\right) +J_{y}^{2}{{\cos }^{2}}\left( {{k_{y}}L}%
\right) +J_{z}^{2}{{\cos }^{2}}\left( {{k_{z}}L}\right) }$ touch at $\left( {%
{k_{x}},{k_{y}},{k_{z}}}\right) =\left( {0,\pm {\pi
\mathord{\left/
 {\vphantom {\pi  {2L, \pm {\pi  \mathord{\left/
 {\vphantom {\pi  {2L}}} \right.
 \kern-\nulldelimiterspace} {2L}}}}} \right.
 \kern-\nulldelimiterspace} {2L, \pm {\pi  \mathord{\left/
 {\vphantom {\pi  {2L}}} \right.
 \kern-\nulldelimiterspace} {2L}}}}}\right) $. Figure \ref{fig:1}(e) depicts
the energy spectra in the BZ, Weyl points and their chiralities. Hamiltonian
$H(\mathbf{k})$ may be compactly written as $H=\mathbf{d}\cdot {\boldsymbol{%
\sigma }}$ , where vector $\mathbf{d}$ has components ${d_{x}}=-2{J_{y}}\cos
\left( {{k_{y}}L}\right) $, ${d_{y}}=-2{J_{x}}\sin \left( {{k_{x}}L}\right) $%
, and ${d_{z}}=2{J_{z}}\cos \left( {{k_{z}}L}\right) $. The Weyl point is a
source of the monopole magnetic field. Here, paired Weyl points with
opposite chiralities may be viewed as a monopole-antimonopole pair in the
momentum space. To show these points, we derive the Berry curvature for the
lowest band \cite{he2012berry}:
\begin{equation}
{F^{a}}={\epsilon _{abc}}{F_{bc}}={\epsilon _{abc}}\left[ {\frac{1}{{2{d^{3}}%
}}\mathbf{d}\cdot \left( {\frac{{\partial \mathbf{d}}}{{\partial {k_{b}}}}%
\times \frac{{\partial \mathbf{d}}}{{\partial {k_{c}}}}}\right) }\right] ,
\label{eq:refname3}
\end{equation}%
where the three components of $\mathbf{F}\left( \mathbf{k}\right) $ are
\begin{equation}
\begin{array}{l}
{F^{x}}={{-8{J_{x}}{J_{y}}{J_{z}}\sin \left( {{k_{x}}L}\right) \sin \left( {{%
k_{y}}L}\right) \sin \left( {{k_{z}}L}\right) }\mathord{\left/ {\vphantom {{
- 8{J_x}{J_y}{J_z}\sin \left( {{k_x}L} \right)\sin \left( {{k_y}L}
\right)\sin \left( {{k_z}L} \right)} {D\left( \mathbf{k} \right)}}} \right.
\kern-\nulldelimiterspace} {D\left( \mathbf{k} \right)}} \\
{F^{y}}={{8{J_{x}}{J_{y}}{J_{z}}\cos \left( {{k_{x}}L}\right) \cos \left( {{%
k_{y}}L}\right) \sin \left( {{k_{z}}L}\right) }\mathord{\left/ {\vphantom
{{8{J_x}{J_y}{J_z}\cos \left( {{k_x}L} \right)\cos \left( {{k_y}L}
\right)\sin \left( {{k_z}L} \right)} {D\left( \mathbf{k} \right)}}} \right.
\kern-\nulldelimiterspace} {D\left( \mathbf{k} \right)}} \\
{F^{z}}={{8{J_{x}}{J_{y}}{J_{z}}\cos \left( {{k_{x}}L}\right) \sin \left( {{%
k_{y}}L}\right) \cos \left( {{k_{z}}L}\right) }\mathord{\left/ {\vphantom
{{8{J_x}{J_y}{J_z}\cos \left( {{k_x}L} \right)\sin \left( {{k_y}L}
\right)\cos \left( {{k_z}L} \right)} {D\left( \mathbf{k} \right)}}} \right.
\kern-\nulldelimiterspace} {D\left( \mathbf{k} \right)}}%
\end{array}%
,  \label{eq:refname4}
\end{equation}%
where \resizebox{0.9\hsize}{!}{$D\left( \mathbf{k}\right) ={\left[ {4J_{x}^{2}{{\sin }^{2}}\left( {{%
k_{x}}L}\right) +4J_{y}^{2}{{\cos }^{2}}\left( {{k_{y}}L}\right) +4J_{z}^{2}{%
{\cos }^{2}}\left( {{k_{z}}L}\right) }\right] ^{{3%
\mathord{\left/
 {\vphantom {3 2}} \right.
 \kern-\nulldelimiterspace}2}}}$}. In Fig. \ref{fig:1}(f), arrows show that
the flux of the Berry curvature $\mathbf{F}\left( \mathbf{k}\right) $ flows
from one monopole to the other, thus defining nontrivial topological
properties of a topological semimetal, where the Weyl points behave as a
sink and source.

The dispersions around the Weyl points are locally linear and described by $%
H\left( \mathbf{q}\right) =\sum\limits_{i,j=\left[ {x,y,z}\right] }{{v_{ij}}{%
q_{i}}}{\sigma _{j}}$, where $\mathbf{q}$ is the displacement momentum with
respect to the momentum of the node, $v_{ij}$ are elements of a $3\times 3$
matrix
\begin{equation}
\left( {%
\begin{array}{ccccccccccccccccccc}
0 & {-2{J_{x}}L} & 0 &     \\
{\pm 2{J_{y}}L} & 0 & 0    \\
0 & 0 & {\pm 2{J_{z}}L}
\end{array}%
}\right) ,\   \label{eq:refname5}
\end{equation}%
and the chirality, which determines the Weyl points, may be defined as $\chi%
\mathrm{=sign}\left( {\det \left[ {{v_{i,j}}}\right] }\right) $.

\begin{figure}[tph]
\centering
{\includegraphics[width=1\linewidth]{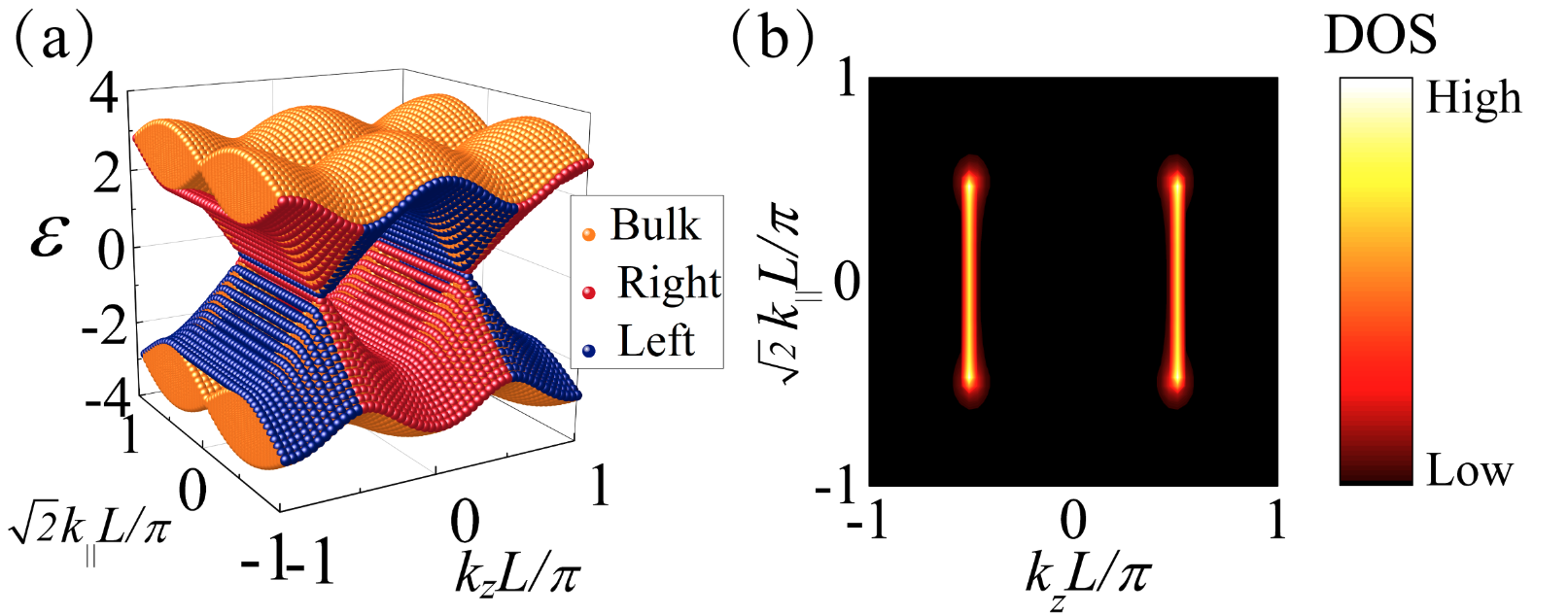}}
\caption{(a) The energy spectrum of the slab, $\protect\varepsilon \left( {{%
k_{\parallel }},{k_{z}}}\right) $, with open boundary conditions. The Weyl
points are connected with Fermi arcs in the momentum space. Two sheets of
surface states, corresponding to the two surface states localized at the
right and left sides of the slab, are Fermi arcs shown by red and blue,
respectively. (b) Zero-energy momentum spectrum of surface states showing the trajectory of the surface Fermi arc,
representing arcs connecting Weyl points with opposite chiralities and the
intersections of the two dispersion sheets.}
\label{fig:2}
\end{figure}

Weyl semimetals imply the existence of topological surface states in the
form of Fermi arcs in the momentum space. The appearance of Fermi arcs is
expected whenever the projections of the Weyl points onto the surface of the
cut (open boundary condition) do not coincide.
Furthermore, the states are highly localized on the surface, and their group
velocity will show a specific propagation direction for each surface. To
observe the surface state induced by the topological properties of the Fermi
arcs, we consider a material maintaining translational
invariance in all the three directions. The bulk-boundary correspondence
suggests that there exist topological surface modes propagating along the
interface of the lattice. We take a slab cut orthogonally to the $\hat{x%
}-\hat{y}$ direction (infinite along the $\hat{z}$ and $\hat{x}+\hat{y}$
directions) in. The slab is considered as a 2D Bravais lattice possessing each good quantum
 number along $k_{\parallel }$ or $k_z$ .The unit
vectors of the Bravais lattice of the slab are ${\mathbf{a}_{1}}=L({\hat{x}+%
\hat{y}})$ and ${\mathbf{a}_{2}}=L\hat{z}$. A generic $\mathbf{k}$-point in
the BZ is expressed as $\mathbf{k}\mathrm{=}{{{k_{\parallel }}\left( {\hat{x}%
+\hat{y}}\right) }%
\mathord{\left/
 {\vphantom {{{k_\parallel }\left( {\hat x + \hat y} \right)} {\sqrt 2 }}} \right.
 \kern-\nulldelimiterspace} {\sqrt 2 }}+{k_{z}}\hat{z}$. Thus, the
projection of Weyl points onto the slab surface are at $\left( {{%
k_{\parallel }},{k_{z}}}\right) =\left( {\pm {\pi
\mathord{\left/ {\vphantom
{\pi  {\sqrt 2 L, \pm {\pi  \mathord{\left/ {\vphantom {\pi  L}} \right.
\kern-\nulldelimiterspace} L}}}} \right. \kern-\nulldelimiterspace} {\sqrt 2
L, \pm {\pi  \mathord{\left/ {\vphantom {\pi  L}} \right.
\kern-\nulldelimiterspace} L}}}}\right) $, and $\varepsilon \left( {{%
k_{\parallel }},{k_{z}}}\right) $ is plotted in Fig. \ref{fig:2}(a). The red
and blue modes are surfaces states localized, respectively, at the right
(RT) and left (LT) edges of the lattice. Specifically, the zero-energy
momentum spectrum $\left( {\varepsilon \mathrm{=}0}\right) $ is calculated
using Eq. \ref{eq:refname6}, which is constructed in terms of the exact
eigenstates of Hamiltonian \cite{ref23,ref24}:
\begin{equation}
\rho \left( {\varepsilon ,\mathbf{k}}\right) =-\frac{1}{\pi }{\mathop{\rm Im}%
\nolimits}Tr\left[ {\frac{1}{{\varepsilon -H\left( \mathbf{k}\right) +i{0^{+}%
}}}}\right] .\   \label{eq:refname6}
\end{equation}%
Figure \ref{fig:2}(b) shows two open-line segments connecting four projected
Weyl nodes with four opposite chiralities, implying the existence of two
separate surface Fermi arcs. To detect the Weyl points, the linear spectra
along the three directions can be measured by means of the momentum-resolved
radio-frequency spectroscopy, which has been utilized for the observation of
the Dirac cone in atomic gases \cite%
{huang2016experimental,meng2016experimental}.

\section{The nonlinear model}

\label{sec:3}

\begin{figure}[htp]
\centering
{\includegraphics[width=1\linewidth]{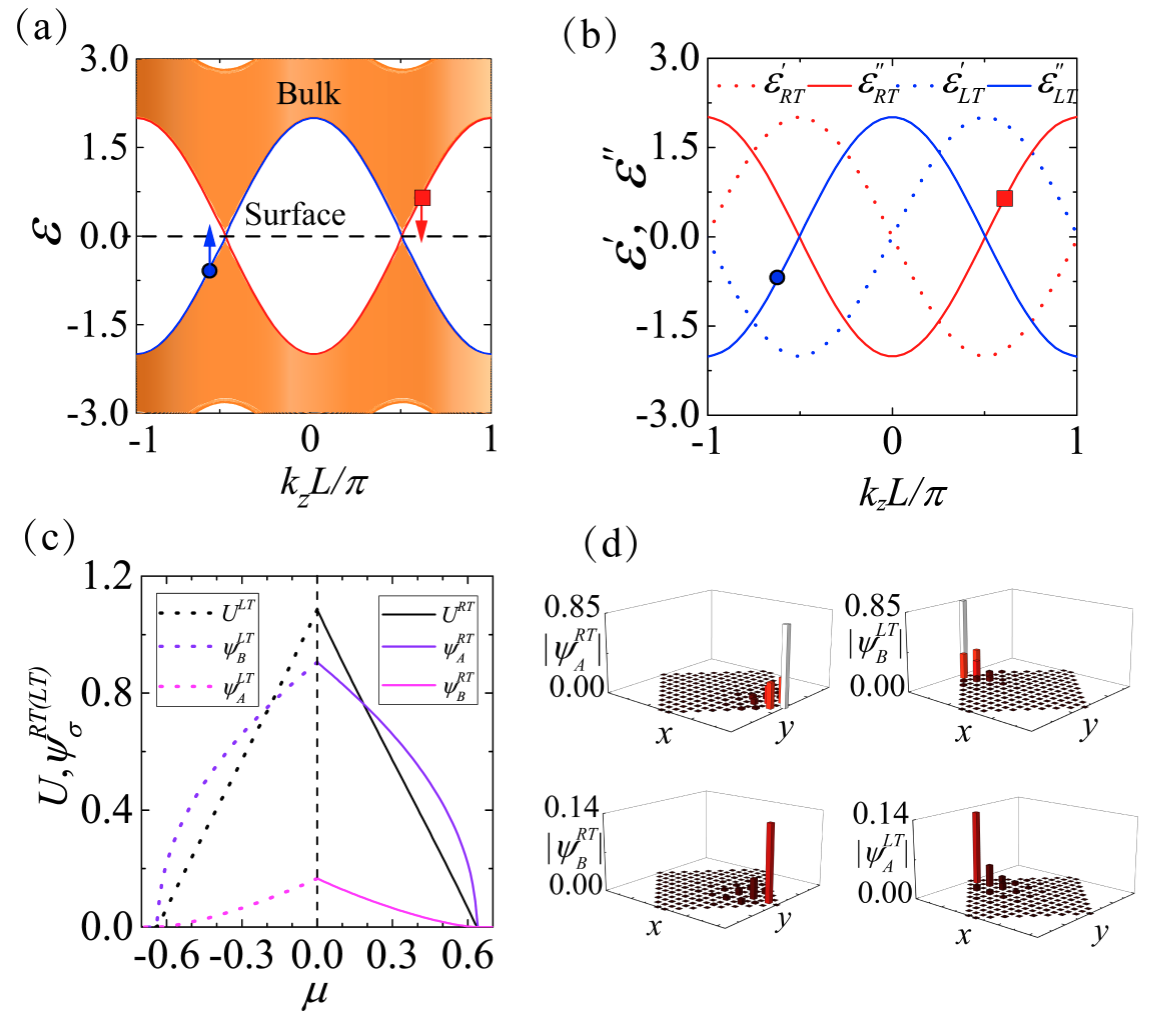}}
\caption{(a) Orange parts correspond to the modes residing in the bulk of
the lattice, while red and blue lines indicate surface states belonging to
different branches. (b) First- and second-order derivatives of the energy of
linear surface states versus momentum ${k_{z}}$. Red square and blue circle denote
parameters corresponding to the modes at ${k_{z}}\mathrm{\ =}-0.4%
\protect\pi ,0.6\protect\pi $. (c) Norm $U_{\protect\sigma }$ per $z$-period
and peak amplitudes $\protect\psi _{\protect\sigma }$ of pseudospin
components $\protect\sigma $ versus $\protect\mu $ for the nonlinear surface
states. The power and amplitude of the nonlinear modes bifurcate from the
zero-intensity point for ${k_{z}}\mathrm{\ =}-0.4\protect\pi ,0.6\protect\pi
$. Solid and dotted curves denote the amplitude of ${\protect\psi _{\protect%
\sigma }^{RT(LT)}}$ component. (d) The transverse profile of $A$ and $B$
components per $z$-period of the nonlinear surface state at right and left
corner of the lattice. }
\label{fig:3}
\end{figure}

Now, we focus on the transport characteristics of the surface mode in the
present system. The time evolution can be described by a scaled system of
coupled Eq. for the $A$ and $B$ components of the spinor wave function $%
\mathbf{\Psi }\mathrm{\ =}{\left( {\psi _{m,n,l}^{A},\psi _{m,n,l}^{B}}%
\right) ^{T}}$:
\begin{equation}
i\frac{{d\psi _{m,n,l}^{\sigma }}}{{dt}}=\sum\limits_{\tau }{{H^{\sigma
,\tau }}\psi _{m,n,l}^{\tau }}.\   \label{eq:refname7}
\end{equation}%
 To introduce the bulk-edge correspondence , we first address the
spectrum of linear modes that are periodic along the $z$-axis and cut
in the $\hat{x}+\hat{y}$ and $\hat{x}-\hat{y}$  direction near the edge. These modes are Bloch functions $\psi
_{m,n,l}^{\sigma }\mathrm{\ =}u_{m,n,l}^{\sigma }{e^{i\varepsilon t+i{k_{z}}%
lL}}$, where ${k_{z}}$ is the Bloch momentum, $\varepsilon $ is the energy
eigenvalue, and the corresponding momentum width of the BZ is given by ${{%
2\pi }%
\mathord{\left/
 {\vphantom {{2\pi } L}} \right.
 \kern-\nulldelimiterspace}L}$. Based on the similarity to multilayer
structures which realize the 3D Weyl-semimetal phase \cite{burkov2011weyl},
we treat the Weyl lattice as a set of identical plane layers. The periodicity
along the $z$ direction insures that $k_{z}$ as an appropriate quantum
number. For each fixed $k_{z}$, the 3D system can be reduced to an effective
2D one with a unit cell in the $(x,y)$ plane \cite{xiao2015synthetic}. The
respective 2D Hamiltonian, $H_{{k_{z}}}^{2D}$, is parameterized by $k_{z}$. The nonzero Chern
number $\pm 2$ of $H_{{k_{z}}}^{2D}$ implies the existence of edge states at
the boundary of finite systems. On the 2D BZ parallel to $k_{z}$,
the chiral surface states wrap around the full
2D BZ forming Fermi arcs \cite%
{kim2015dirac,mullen2015line,lim2017pseudospin}. Utilizing the 2D
equivalence, stationary states of Eqs. (\ref{eq:refname7}) under open
boundary condition corresponding to $H_{{k_{z}}}$ are denoted as
\begin{equation} \label{eq:refname8}
\resizebox{1\hsize}{!}{%
$ \sum\limits_{\tau }{\left( {{e^{-i{k_{z}}L}}H_{m,n,l\mathrm{\ -}1}^{\sigma
,\tau }\mathrm{\ +}H_{m,n,l}^{\sigma ,\tau }\mathrm{\ +}{e^{i{k_{z}}L}}%
H_{m,n,l\mathrm{\ +}1}^{\sigma ,\tau }}\right) }u_{m,n,l}^{\tau
}=\varepsilon u_{m,n,l}^{\sigma }.\ $%
        }
\end{equation}
Here, indices $m,n$ enumerate the unit cells in the $x$ and $y$ directions,
and $l$ denotes a specific layer along the $z$-axis. For a single atom,
energy eigenstates are Bloch wave functions, or an appropriate superposition
of Bloch states which are well localized on individual lattice sites. To
simplify the notation and relate the indices to the coordinates, $\psi
_{m,n,l}^{\sigma }\mathrm{\ =}u_{m,n,l}^{\sigma }{e^{i\varepsilon t+i{k_{z}}%
lL}}$ can be rewritten as ${\psi _{\sigma }}\left( {x,y,z}\right) ={%
e^{i\varepsilon t+i{k_{z}}z}}{u_{\sigma }}\left( {x,y,z}\right) $. A
representative spectrum for the lattice with the surface states is shown in
Fig. \ref{fig:3}(a) in the form of the energy-momentum diagrams for the $%
\left[ {{{\ -\pi }%
\mathord{\left/
 {\vphantom {{ - \pi } {L{\pi  \mathord{\left/
 {\vphantom {\pi  L}} \right.
 \kern-\nulldelimiterspace} L}}}} \right.
 \kern-\nulldelimiterspace} {L,{\pi \mathord{\left/
 {\vphantom {\pi  L}} \right.
 \kern-\nulldelimiterspace} L}}}}\right] $ interval by solving Eqs. (\ref%
{eq:refname8}). Due to the spinor character of the model, the spectrum
consists of two groups of bands. The spectrum shows two Weyl points at $\pm {%
{{k_{z}}L}%
\mathord{\left/
 {\vphantom {{{k_z}L} {2\pi }}} \right.
 \kern-\nulldelimiterspace} {2\pi }}$, where the upper and lower bands touch
each other.
In Fig. \ref{fig:3}(b), we plot the dispersion coefficient $\varepsilon
^{\prime \prime }$ as a function of ${k_{z}}$ corresponding to different
linear surface states. RT (LT) branches in the momentum intervals presenting
the localized states near the right (left) edge are denoted by the red
(blue) color. All such intervals for every linear surface mode give rise to
surface solitons which are investigated below. For every branch, there exist
two unique ${k_{z}}$ values where dispersion $\varepsilon ^{\prime \prime }$
vanishes and wave packets with a broad envelope may evolve almost without
broadening even in the linear limit, due to the vanishing of the dispersion
near the Weyl points.

To study dynamics of the nonlinear surface excitations and capture transport
characteristics in the nonlinear model, nonlinear terms are added
to Eqs. (\ref{eq:refname7}). Hence, the nonlinear evolution of the vector
wave function with nonlinear interaction is governed by the following
Eqs.:
\begin{equation}
i\frac{{d\psi _{m,n,l}^{\sigma }}}{{dt}}=\sum\limits_{\tau }{{H^{\sigma
,\tau }}\psi _{m,n,l}^{\tau }}+{N^{\sigma }}\left( {\psi _{m,n,l}^{\sigma }}%
\right) \psi _{m,n,l}^{\sigma },\   \label{eq:refname9}
\end{equation}%
where $H_{\sigma }$ maintains the form of the coupling matrix of the
lattice, and ${N^{\sigma }}\left( {\psi _{m,n,l}^{\sigma }}\right) $
represents the diagonal matrix with nonlinear elements \cite%
{ref25,ref26,ref27}:

\begin{equation}
\left[ {{N^\sigma }\left( {\psi _{m,n,l}^\sigma } \right)} \right] = {%
g^\sigma }\left( {{{\left| {\psi _{m,n,l}^\sigma } \right|}^2}} \right).\
\label{eq:refname10}
\end{equation}
Here, self-interaction nonlinearity can be achieved by applying the mean-field
theory (variables $\psi_{\sigma}$ represent large clusters trapped at different sites
of the lattice, rather than individual atoms). Although the Pauli principle
does not allow the direct self-interaction, an effective self-interaction may
be induced via the local-field effect, i.e., local deformation of the optical
lattice by the atomic gas \cite{li2008matter,zhu2011strong,dong2013polaritonic}.
By applying the continuum approximation \cite{christodoulides1988discrete,
hudock2004anisotropic}(${k_{z}}d\ll 1$ , $d$ is the step size of the Taylor
expansion), the diffraction coefficients along the $z$-axis of the two
components are denoted as $\gamma _{z}^{\sigma }=\pm {d^{2}}\cos \left( {{%
k_{z}}L}\right) $. Under this approximation, the type of diffraction can be
compensated by the nonlinearity.
For given $k_{z}$, the nonlinear coefficient ${g^{\sigma }}$ in Eqs. (\ref{eq:refname10}%
) is rescaled to be ${g^{\sigma }}\mathrm{\ =}\pm 1$ for repulsive and
attractive interatomic interactions, respectively \cite{ref28}. Nonlinear
solutions are introduced in the same form as the linear ones,
\begin{equation}
\psi _{m,n,l}^{\sigma }\mathrm{\ =}\phi _{m,n,l}^{\sigma }{e^{i\mu t+i{k_{z}}%
lL}},  \label{psi}
\end{equation}%
which may also be written as: ${\psi _{\sigma }}\left( {x,y,z}\right) ={%
u_{\sigma }}\left( {x,y,z}\right) {e^{i\mu t+i{k_{z}}z}}$. The substitution
of ansatz (\ref{psi}) in Eqs. (\ref{eq:refname9}) leads to the stationary
version of the nonlinear Eqs.:
\begin{equation}
\mu \phi _{m,n,l}^{\sigma }=\sum\limits_{\tau }{{H^{\sigma ,\tau }}\phi
_{m,n,l}^{\tau }}+{N^{\sigma }}\left( {\phi _{m,n,l}^{\sigma }}\right) \phi
_{m,n,l}^{\sigma },\   \label{phi}
\end{equation}%
Because the nonlinearity in our model dominates over the interaction
between the pseudospin components, nonlinear solutions exist when
nonlinearity-induced energy eigenvalue, $\mu $, does not fall into the bulk
band (i.e., it belongs to the spectral gap), and they vanish for $\mu $
approaching the linear limit, $\varepsilon $.
We numerically solved Eqs. (\ref{phi}), using the Newton's method in the
frequency domain. In Fig. \ref%
{fig:3}(c), the consideration of the vicinity of the zero point suggests
that, for the same displacement of $k_{z}$ away from the Weyl points, the
right- and left-side modes possess the same characteristics except for the
fact that soliton solutions emerge at opposite signs of the nonlinearity.
 The nonlinear surface states are characterized by dependence
of the total norm $U\mathrm{\ =}{U_{A}}+{U_{B}}$ of the $A$ and $B$ wave
component per $z$-period on $\mu $, where ${U_{\sigma }}=\int_{-L/2}^{L/2}{dz%
{{\iint \left\vert {\psi _{\sigma }(x,y)}\right\vert }^{2}}dxdy}$ are norms
of the two components per $z$-period. The evidence that nonlinear states
bifurcate from the linear ones is provided by dependencies of the peak
amplitudes, $\max \left\vert {\psi _{\sigma }}\right\vert $, on $\mu $. The
vanishing of $\max \left\vert {\psi _{\sigma }}\right\vert $ at the
bifurcation point indicates the thresholdless character of the nonlinear
surface states. In addition, amplitude profiles of nonlinear modes $\psi
_{\sigma }^{\mathrm{RT}(\mathrm{LT})}$ for different types of the
nonlinearity, localized at the $A$-$B$ site sets are shown in Fig. \ref%
{fig:3}(d). It is seen that the modes are almost identical, except for being
localized at different corner of the lattice. Therefore, it is sufficient to
analyze the single species of the modes. Thus we suppress the superscript
RT(LT), and focus on the mode attached to right corner. The existence interval
of $\mu $ for nonlinear modes is determined by energy difference $\delta
=\varepsilon -\mu $ between the linear and nonlinear state for given $k_{z}$%
. This difference, representing the energy separation from the bulk modes,
means, as mentioned above, that the nonlinear states, localized along the $z$%
-axis, may only exist with $\mu $ falling into a gap of the spectral
structure. When $\mu $ crosses the edge of the spectral band, the nonlinear
mode will lose the localization and couple with the bulk modes (embedded
solitons, which may exist, as exceptional states, in Bloch bands of some
discrete nonlinear systems \cite{yagasaki2005discrete}, were not found here).

\begin{figure}[htb]
\centering
{\includegraphics[width=1\linewidth]{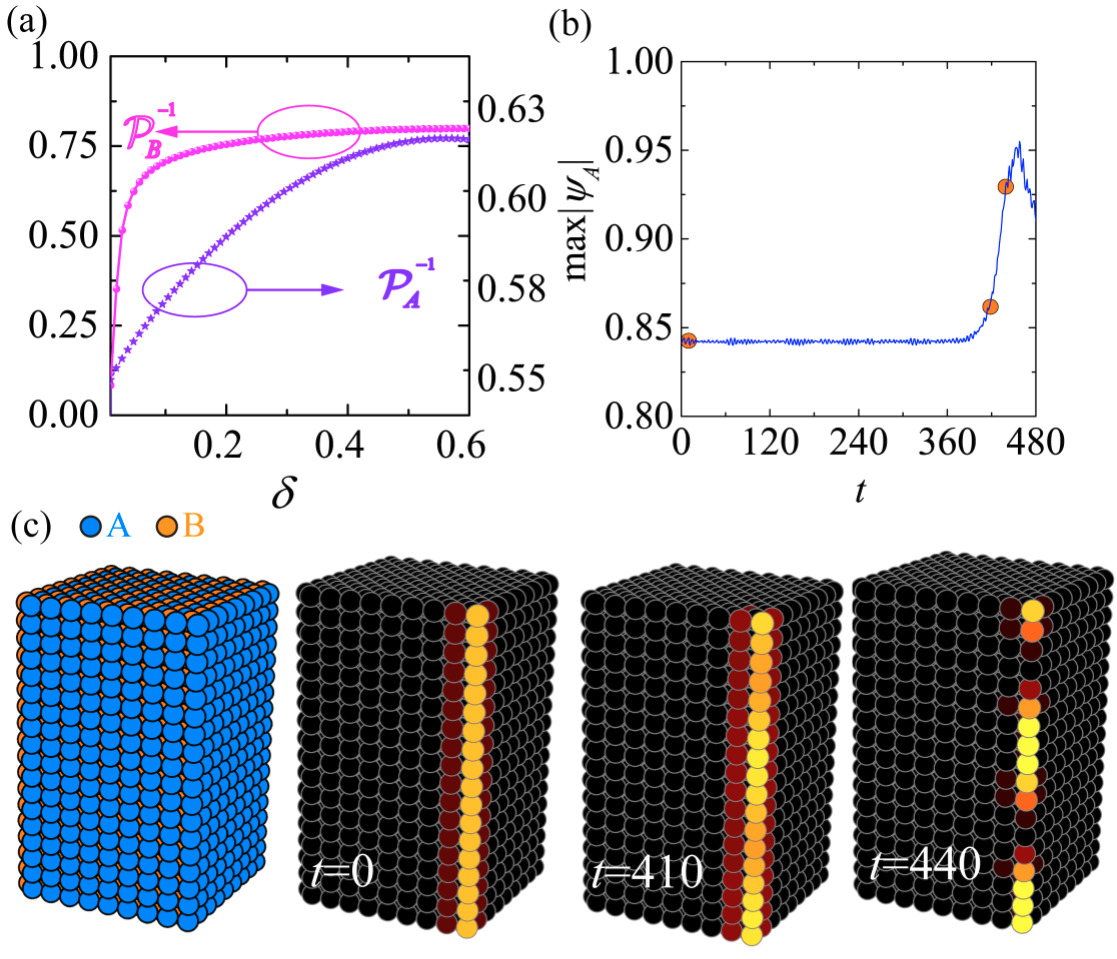}}
\caption{(a) The inverse participation number $\mathcal{P}%
_{\sigma }^{\mathrm{\ -}1}$ versus energy difference $\protect\delta $ corresponding to
different sites. (b) The stable evolution of perturbed nonlinear surface
states corresponding to $\protect\mu \mathrm{\ =}0.1$ and ${k_{z}}\mathrm{\ =%
}0.6\protect\pi $. The evolution of peak amplitudes $\max \left\vert {%
\protect\psi _{A}}\right\vert $ in $t$ testifies to the stability of the
nonlinear wavepacket. (c) The distribution of the $A$ and $B$ sites and the
evolution pattern of $\protect\psi _{\protect\sigma }$ versus $t$, displayed
in the 3D form.}
\label{fig:4}
\end{figure}

To quantify the soliton's localization, we use the inverse participation
number:
\begin{equation} \label{eq:refname11}
\resizebox{1\hsize}{!}{%
$ {\mathcal{P}_{\sigma }^{\mathrm{\ -}1}}={{\int_{-L/2}^{L/2}{dz\iint {{{%
\left\vert \psi _{\sigma }\right\vert }^{4}}}dxdy}}\mathord{\left/
{\vphantom {{\int_{ - L/2}^{L/2} {dz\iint {{{\left| \psi _\sigma
\right|}^4}} dxdy} } {{{\left( {\int_{ - L/2}^{L/2} {dz\iint {{{\left| \psi
_\sigma \right|}^2}} dxdy} } \right)}^2}}}} \right.
\kern-\nulldelimiterspace} {{{\left( {\int_{ - L/2}^{L/2} {dz\iint {{{\left|
\psi _\sigma \right|}^2}} dxdy} } \right)}^2}}}.\  $%
}
\end{equation}
Figure \ref{fig:4}(a) plots $\delta $ versus the inverse participation
number $\mathcal{P}_{\sigma }^{\mathrm{\ -}1}$, the colored lines pertaining
to different sites $A$ and $B$. With the increasing of $\delta $, the
localization monotonically strengthens with the increase of $\mathcal{P}%
_{\sigma }^{\mathrm{\ -}1}$.

Figure \ref{fig:4}(b) reports results of the stability analysis for the
nonlinear surface states, performed by perturbing them with a small
broadband input noise ($1\%$ in amplitude), and continuously tracing their
subsequent evolution up to very large times.  It is seen that the perturbed nonlinear modes maintain themselves even at $t>{360}$. Eventually, the nonlinear surface states are
unstable due to the modulation instability in the periodic potential \cite%
{meier2004experimental,lederer2008discrete}. However, rather than decaying,
the $z$-periodic periodic surface state breaks up into soliton trains (see
the pattern at $t=440$ in Fig. \ref{fig:4}(c)). These results suggest that
the nonlinearity may indeed build robust surface Weyl solitons%
, bifurcating from linear surface states in the Weyl lattice.
\begin{figure}[htp]
\centering
{\includegraphics[width=1\linewidth]{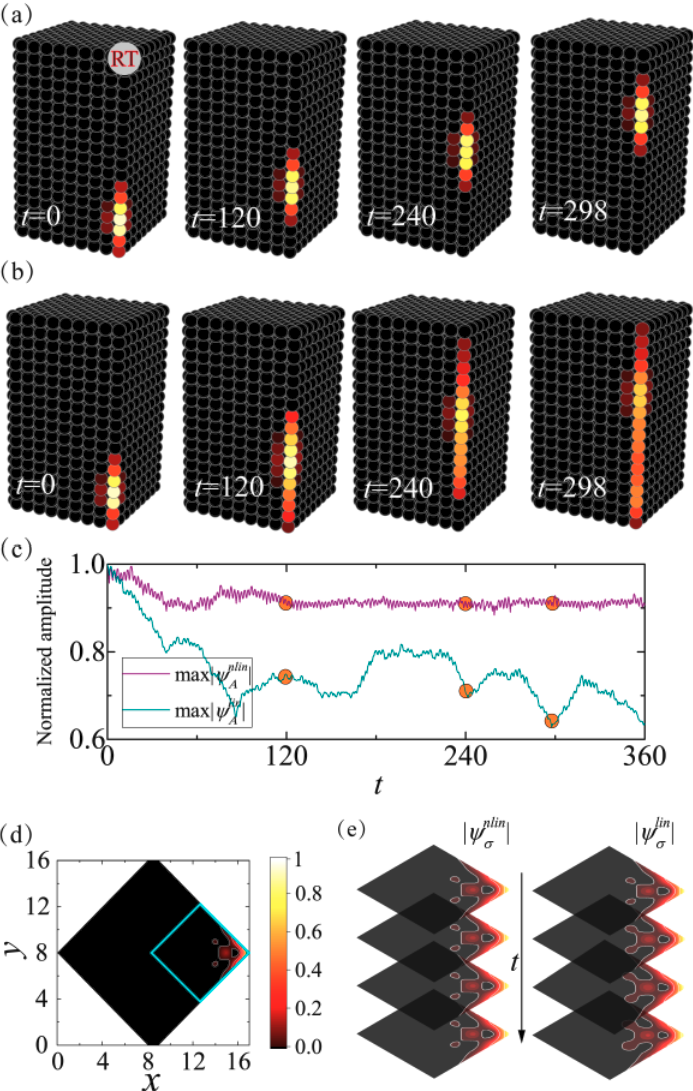}}
\caption{The propagation of a nonlinear wavepacket at $\protect\delta =0.12$%
. To stress the fact that the surface state moves along the $z$-direction,
and to provide details of its shape, we show distributions of $\protect\psi %
_{A}$ in the 3D window at times $t=0;120;240;298$. (a) At the initial
moment, $t=0$, the input beam is localized at the surface with ${k_{z}}%
\mathrm{\ =}0.6\protect\pi $. The wavepacket moves along the $z$-axis on the
left or right side of the bulk lattice. The shape of the wavepacket is
preserved, under the action of the nonlinearity. (b) With the same input,
the wavepacket is deformed in the course of the evolution without the
nonlinearity. (c) The dependence of the peak amplitude of the $\protect\psi %
_{A}$ component on $t$ for the linear and nonlinear systems, with the orange
circles representing the time moments used in Figs. \protect\ref{fig:5}(a)
and (b). (d) The panel shows normalized isocontours of the density of $%
\left\vert {\protect\psi _{\protect\sigma }}\right\vert $. (e) Zoomed
evolution in the cyan square in Fig. \protect\ref{fig:5}(d), with or without
the nonlinearity.}
\label{fig:5}
\end{figure}
To develop a more regular approach for demonstrating the existence of the
Weyl solitons, we rewrite Eqs. \ref{eq:refname9} as $i\partial \mathbf{\Psi }%
/\partial t=\mathcal{L}\mathbf{\Psi }+\mathcal{N}\mathbf{\Psi }$, where $%
\mathbf{\Psi }\mathrm{\ =}{\left( {{\psi _{A}},{\psi _{B}}}\right) ^{T}}$,
operator $\mathcal{L}\mathrm{\ =}H$ includes all linear terms, while
operator $\mathcal{N}$ accounts for the nonlinearity. Then, the expression of the soliton can
be written as \cite{ablowitz2011nonlinear}
\begin{equation} \label{eq:refname12}
\resizebox{1\hsize}{!}{%
 $ \begin{array}{l}
\mathbf{\mathbf{\Psi }}\left( {x,y,z,t}\right) =\sum\limits_{j}{\int_{-{\pi %
\mathord{\left/ {\vphantom {\pi L}} \right. \kern-\nulldelimiterspace}L}}^{{%
\pi \mathord{\left/ {\vphantom {\pi L}} \right. \kern-\nulldelimiterspace}L}}%
{a_{j}}\left( {\kappa ,t}\right) \mathbf{u}\left( {x,y,z,{k_{z}}+\kappa }%
\right) {e^{i\varepsilon t+i\left( {{k_{z}}+\kappa }\right) z}}d\kappa } \\
\approx \int_{-{\pi \mathord{\left/ {\vphantom {\pi L}} \right.
\kern-\nulldelimiterspace}L}}^{{\pi \mathord{\left/ {\vphantom {\pi L}}
\right. \kern-\nulldelimiterspace}L}}a\left( {\kappa ,t}\right) \mathbf{u}%
\left( {x,y,z,k+\kappa }\right) {e^{i\varepsilon t+i\left( {{k_{z}}+\kappa }%
\right) z}}d\kappa ,%
\end{array} $%
}
\end{equation}
where vector function $\mathbf{u}={\left( {{u_{A}},{u_{B}}}\right) ^{T}}$
satisfies the linear Eq. $\left( {\mathcal{L}+\varepsilon }\right)
\mathbf{u}{e^{i{k_{z}}z}}=0$ for the linear Bloch mode with momentum ${k_{z}}
$, and we take into account that the corresponding energy, $\varepsilon $,
depends on quasi-momentum ${k_{z}}$. Here $\kappa $ is the momentum offset
from the carrier soliton momentum ${k_{z}}$, and amplitude $a\left( {\kappa
,t}\right) $ is assumed to be well localized in $\kappa $. Using the Taylor
expansion in $\kappa $ for $\mathbf{u}\left( {x,y,z,{k_{z}}+\kappa }\right) $
in the above integral, one obtains the expression for the shape of the
surface-state wave packet:
\begin{equation}
\mathbf{\Psi }\left( {x,y,z,t}\right) ={e^{i\varepsilon t+i{k_{z}}z}}%
\sum\limits_{j=0,\infty }{\frac{{{{\left( {\ -i}\right) }^{j}}}}{{j!}}}\frac{%
{{\partial ^{j}}\mathbf{u}}}{{\partial {k_{z}}^{j}}}\left[ {\frac{{{\partial
^{j}}a\left( {z,t}\right) }}{{\partial {z^{j}}}}}\right] ,\
\label{eq:refname13}
\end{equation}%
where $a\left( {z,t}\right) =\int_{-{\pi
\mathord{\left/
 {\vphantom {\pi  L}} \right.
 \kern-\nulldelimiterspace}L}}^{{\pi
\mathord{\left/
 {\vphantom {\pi  L}} \right.
 \kern-\nulldelimiterspace}L}}a\left( {\kappa ,t}\right) {e^{i{k_{z}}z}}%
d\kappa $ is the envelope function of the corresponding nonlinear surface
state. To see how $\mathcal{L}$ acts on wave function $\mathbf{\Psi }$, we
move $\mathcal{L}$ through the integral and take $\mathcal{L}\mathbf{u}{e^{i{%
k_{z}}z}}=\mathrm{\ -}\varepsilon \mathbf{u}{e^{i{k_{z}}z}}$ into account,
arriving at
\begin{equation}\label{eq:refname14}
\resizebox{1\hsize}{!}{%
$\mathcal{L}\mathbf{\Psi }=-\int_{-{\pi \mathord{\left/ {\vphantom {\pi L}}
\right. \kern-\nulldelimiterspace}L}}^{{\pi \mathord{\left/ {\vphantom {\pi
L}} \right. \kern-\nulldelimiterspace}L}}{\varepsilon \left( {{k_{z}}+\kappa
}\right) a}\left( {\kappa ,t}\right) \mathbf{u}\left( {x,y,z,{k_{z}}+\kappa }%
\right) {e^{i\varepsilon t+i\left( {{k_{z}}+\kappa }\right) z}}d\kappa .\ $%
}
\end{equation}
Employing the Taylor series expansion in $\kappa $ for both $\varepsilon
\left( {{k_{z}}+\kappa }\right) $ and $\mathbf{u}\left( {x,y,z,{k_{z}}%
+\kappa }\right) $, we further obtains
\begin{equation}
\mathcal{L}\mathbf{\Psi }=-{e^{i\varepsilon t+ik_{z}z}}\sum\limits_{j=0,%
\infty }{\frac{{{{\left( {\ -i}\right) }^{j}}}}{{j!}}}\frac{{{\partial ^{j}}%
\left( {\varepsilon \mathbf{u}}\right) }}{{\partial {k_{z}}^{j}}}\left[ {%
\frac{{{\partial ^{j}}{a\left( {z,t}\right) }}}{{\partial {z^{j}}}}}\right]
.\   \label{eq:refname15}
\end{equation}

Assuming that $\mathbf{u}$ changes with ${k_{z}}$ much slower than
eigenvalue $\varepsilon $, the slowly-varying-amplitude approximation
results in the envelope of a forward-travelling wave, slowly varying in time
and space compared to the underlying period, and the underlying solution can
be obtained in the approximate form, eliminating terms with higher-order
partial derivatives. This approximation allows us to keep only the $j=0$
term in Eqs. \ref{eq:refname13}, so that $\mathbf{\Psi }\left( {x,y,z,t}%
\right) ={e^{i\varepsilon t+i{k_{z}}z}}\mathbf{u}\left( {x,y,z,t}\right)
a\left( {z,t}\right) $ and ${{{\partial ^{j}}\left( {\varepsilon \mathbf{u}}%
\right) }%
\mathord{\left/
 {\vphantom {{{\partial ^j}\left( {\varepsilon \mathbf{u}} \right)} {\partial {k_z}^j}}} \right.
 \kern-\nulldelimiterspace} {\partial {k_z}^j}}\approx \mathbf{u}{{{\partial
^{j}}\varepsilon }%
\mathord{\left/
 {\vphantom {{{\partial ^j}\varepsilon } {\partial {k_z}^j}}} \right.
 \kern-\nulldelimiterspace} {\partial {k_z}^j}}$, while the nonlinear term $%
\mathcal{N}$ simplifies to $a{\left\vert a\right\vert ^{2}}\mathbf{u}{%
e^{i\varepsilon t+i{k_{z}}z}}$. Finally, we multiply Eq. $i\partial
\mathbf{\Psi }/\partial t=\mathcal{L}\mathbf{\Psi }+\mathcal{N}\mathbf{\Psi }
$ by ${\mathbf{u}^{\dag }}$ and integrate it over one period along the $z$%
-axis and over the entire $\left( x,y\right) $ plane, which allows us to
derive the nonlinear Schr\"{o}dinger Eq. for the envelope
function:
\begin{equation}
i\frac{{\partial a}}{{\partial t}}=i\varepsilon ^{\prime }\frac{{da}}{{dz}}%
\mathrm{\ +}\frac{1}{2}\varepsilon ^{\prime \prime }\frac{{{d^{2}}a}}{{d{%
z^{2}}}}\mathrm{\ +}{g_{\mathrm{eff}}\left\vert a\right\vert ^{2}a}.\
\label{eq:refname16}
\end{equation}%
Here, we keep only the first two terms proportional to $\varepsilon ^{\prime
}={{\partial \varepsilon }%
\mathord{\left/
 {\vphantom {{\partial \varepsilon } {\partial {k_z}}}} \right.
 \kern-\nulldelimiterspace} {\partial {k_z}}}$ and $\varepsilon ^{\prime
\prime }={{{\partial ^{2}}\varepsilon }%
\mathord{\left/
 {\vphantom {{{\partial ^2}\varepsilon } {\partial {k_z}^2}}} \right.
 \kern-\nulldelimiterspace} {\partial {k_z}^2}}$ in the Taylor expansion of $%
\varepsilon \left( {k_{z}}\right) $. The effective nonlinear coefficient is $%
{g_{\mathrm{eff}}}={{\iiint {{\mathbf{u}^{\dag }}\mathcal{N}\mathbf{u}dxdydz}%
}%
\mathord{\left/
 {\vphantom {{\iiint {{\mathbf{u}^\dag }{\cal N}udxdydz} } {\iiint {{\mathbf{u}^\dag }udxdydz} }}} \right.
 \kern-\nulldelimiterspace} {\iiint {{\mathbf{u}^\dag }\mathbf{u}dxdydz} }}$%
. This coefficient can be calculated numerically for different values of ${%
k_{z}}$, using the linear Bloch modes. When ${\varepsilon ^{^{\prime \prime
}}}>0$ [which corresponds to the red square in Fig. \ref{fig:3}(a)], Eq. (%
\ref{eq:refname16}) admits the bright-soliton solution:
\begin{equation}\label{eq:refname17}
 \resizebox{1\hsize}{!}{%
 $ a_{\mathrm{bright}}\left( {z,t}\right) ={\left( {{{2\delta }\mathord{\left/
{\vphantom {{2\delta } {{g_{\rm{eff}}}}}} \right. \kern-\nulldelimiterspace}
{{g_{\rm{eff}}}}}}\right) ^{{1\mathord{\left/ {\vphantom {1 2}} \right.
\kern-\nulldelimiterspace}2}}}{\mathop{\rm sech}\nolimits}\left[ {{{\left( {{%
{2\delta }\mathord{\left/ {\vphantom {{2\delta } {{\varepsilon ^{''}}}}}
\right. \kern-\nulldelimiterspace} {{\varepsilon ^{''}}}}}\right) }^{{1%
\mathord{\left/ {\vphantom {1 2}} \right. \kern-\nulldelimiterspace}2}}}%
\left( {z+{\varepsilon ^{^{\prime }}}t}\right) }\right] {e^{-i\delta t}}.\ $%
}
\end{equation}
Note that the energy shift $\delta $ in Eq. (\ref{eq:refname17}), introduced
by the nonlinearity, leads to the consequence that the total wave function $%
\mathbf{\Psi }\left( {x,y,z}\right) ={e^{i\varepsilon t+i{k_{z}}z}}\mathbf{u}%
\left( {x,y,z,{k_{z}}}\right) a\left( {z,t}\right) $ varies as ${e^{-i\mu t}}
$. The shift serves as a compensation parameter for the energy difference
between the bulk and nonlinear modes, which are used as the complete set of
nonlinear wave functions. Figure \ref{fig:5}(a) shows the evolution of
solitons constructed as per Eqs. (\ref{eq:refname13}) with the envelope
function given by Eq. (\ref{eq:refname17}). At the chosen value of $k_{z}$,
dispersion $\varepsilon ^{\prime \prime }$ and eigenstate $\mathbf{u}$ were
found by numerically solving the eigenvalue problem defined by Eqs. (\ref%
{eq:refname8}). In this context, periodic boundary conditions along $z$ axis were
used.

A Weyl soliton, that starts its evolution being localized at the right side
of the Weyl lattice features unidirectional motion. This can be interpreted
as follows: unidirectional transport of surface modes relies on the global
topology of the lattice, as a consequence of its specific topological protected band structure,
while the localization of the wave
packet is governed by the nonlinearity. One can see that, after an initial
transient period, when the peak amplitude of the input wave form decreases
due to internal reshaping of its profile, the soliton's amplitude of soliton
remains almost constant [see the purple curve in Fig. \ref{fig:5}(d) showing
the evolution of the peak amplitude $\max \left\vert {\psi _{A}}\right\vert $
of the $\psi _{A}$ component], with the velocity which is nearly identical
to $\varepsilon ^{\prime }$. Sets of similar long-lived nonlinear surface
states can be generated, varying the respective value of $k_{z}$. To confirm
that the localized states indeed exist due to the nonlinearity, we used the
same input, while nonlinearity was switched off. Figure \ref{fig:5}(b) shows
snapshots of the dynamics associated with the surface states in the absence
of the nonlinearity. We observe expansion of the wave packet in Fig. \ref%
{fig:5}(b) and decay of its amplitude, as shown by the green curve in Fig. %
\ref{fig:5}(c). In addition, we introduce the 2D cross section of the 3D
domain at the central position of the wave packet in the $z$ direction, and
display the density of $\left\vert {\psi _{\sigma }}\right\vert $ by means
of isocontours in the 2D plane, in Fig. \ref{fig:5}(c), for the same four
time moments that were chosen in Fig. \ref{fig:5}(e). It is observed that
snapshots of the nonlinear mode maintained their shape, while the linear
mode is spreading out.

\begin{figure}[htp]
\centering
{\includegraphics[width=1\linewidth]{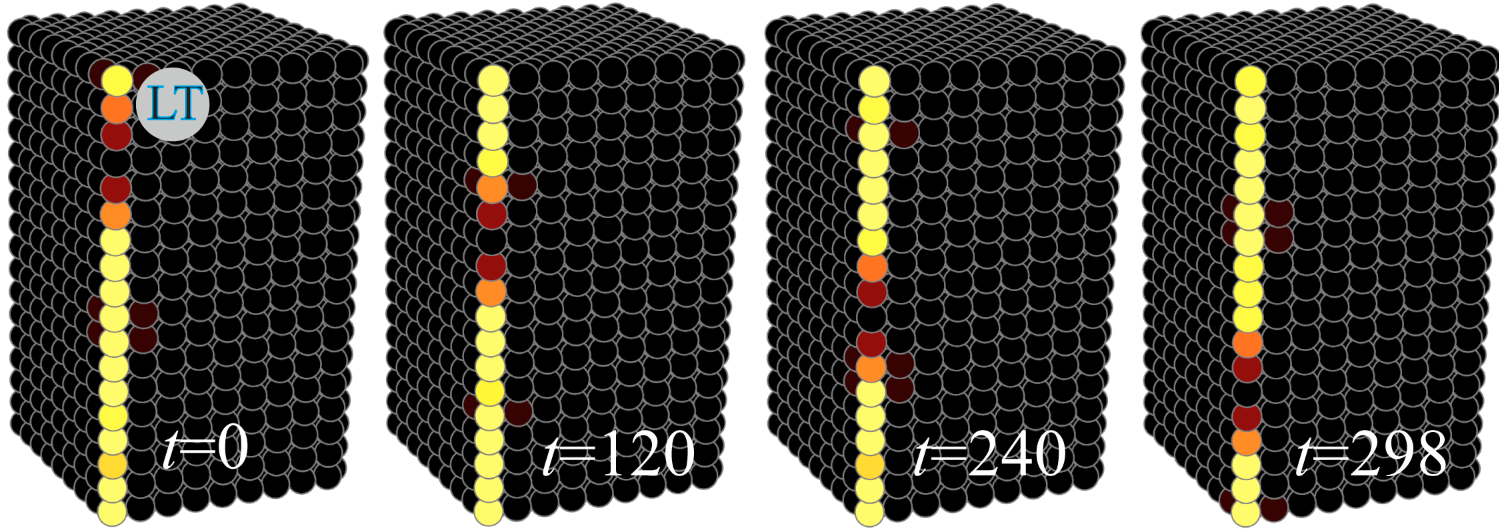}}
\caption{The stable evolution of the dark soliton in the nonlinear system is
shown for $\protect\delta =-0.12$ and ${k_{z}}\mathrm{\ =}-0.4\protect\pi $.
The dark spot moves without any notable deformation. }
\label{fig:6}
\end{figure}
In addition, for ${\varepsilon }^{\prime \prime }<0$ [which corresponds to
the blue circle in Fig. \ref{fig:3}(a)], the system gives rise to dark
solitons, with the envelope:
\begin{equation}\label{eq:refname18}
\resizebox{1\hsize}{!}{%
${a_{\mathrm{dark}}}\left( {z,t}\right) ={\left( {{{\ -2\delta }%
\mathord{\left/ {\vphantom {{ - 2\delta } {{g_{\rm{eff}}}}}} \right.
\kern-\nulldelimiterspace} {{g_{\rm{eff}}}}}}\right) ^{{1\mathord{\left/
{\vphantom {1 2}} \right. \kern-\nulldelimiterspace}2}}}\tanh \left[ {{{%
\left( {{{\ -2\delta }\mathord{\left/ {\vphantom {{ - 2\delta }
{{\varepsilon ^{''}}}}} \right. \kern-\nulldelimiterspace} {{\varepsilon
^{''}}}}}\right) }^{{1\mathord{\left/ {\vphantom {1 2}} \right.
\kern-\nulldelimiterspace}2}}}\left( {z+{\varepsilon ^{^{\prime }}}t}\right)
}\right] {e^{i\delta t}}. $%
}
\end{equation}
In Fig. \ref{fig:6}, we show the evolution of the surface state constructed
using this envelope and Bloch modes $\mathbf{u}$. In the simulations, the
dark soliton also survives for a long time, keeping its initial shape and propagating
along an opposite direction of the bight one.

\section{Conclusion}

The aim of this work is to demonstrate the existence of topological Weyl
surface solitons in the 3D optical lattice. To the best of our knowledge,
effects of the nonlinear were not previously studied in settings emulating
Weyl semimetals by dint of the appropriately designed optical lattice with
an ultracold atomic gas loaded into it.  This proposal also offers a new feasible control of the particles governed by Weyl Eq., which may conquer the difficulty
of physical realization in the real material. Robust modes in the
form of Weyl solitons are revealed by the systematic
analysis of the 3D nonlinear model. Note that the Weyl solitons
arising here should be distinguished from gap solitons. Being formed from
topological surface modes, the Weyl solitons can only propagate along surfaces
of the lattice, in contrast to gap solitons, which propagate in the bulk. The Weyl-soliton
states bifurcate from linear surface modes at zero intensity, indicating the
absence of any threshold necessary for their existence. In addition, bright and dark Weyl soliton perform
an intriguing counter-propagation unidirectional characteristics. Furthermore, the analysis developed in this work is
also applicable to optical waveguides \cite{dreisow2010classical,keil2015optical} and
 nanowires \cite{oreg2010helical,das2012zero} which may support Weyl
solitons, therefore making our results very general and
of relevance to the systems beyond optical lattices. In closing, Weyl solitons,
 the last member of the topological soliton family, may pave the way for the
  realization of many fascinating topological nonlinear phenomena.

\section*{Funding Information}

National Natural Science Foundation of China (NSFC) (61475101); Innovation
Program of Shanghai Municipal Education Commission (13ZZ022); the joint
program in physics between NSF and Binational (US-Israel) Science Foundation
(project No. 2015616); Israel Science Foundation (grant No. 1286/17).

\section*{Acknowledgments}

 Special thanks to Xianfeng Chen and Fangwei Ye for discussion.


\begin{thebibliography}{99}
\newcommand{\enquote}[1]{``#1''}

\bibitem{ref1}
K.~Nomura and A.~H. MacDonald, \enquote{\protect{Quantum transport of massless
  Dirac fermions},} Physical Review Letters \textbf{98}, 076602 (2007).

\bibitem{ref2}
L.~Fu and C.~L. Kane, \enquote{\protect{Probing neutral Majorana fermion edge
  modes with charge transport},} Physical Review Letters \textbf{102}, 216403
  (2009).

\bibitem{ref3}
J.~P. Xu, M.~X. Wang, Z.~L. Liu, J.~F. Ge, X.~Yang, C.~Liu, Z.~A. Xu, D.~Guan,
  C.~L. Gao, D.~Qian, Y.~Liu, Q.~H. Wang, F.~C. Zhang, Q.~K. Xue, and J.~F.
  Jia, \enquote{\protect{Experimental detection of a Majorana mode in the core
  of a magnetic vortex inside a topological insulator-superconductor
  $\rm Bi_2Te_3$/ $\rm NbSe_2$ heterostructure},} Physical Review Letters \textbf{114},
  017001 (2015).

\bibitem{ref4}
S.~M. Huang, S.~Y. Xu, I.~Belopolski, C.~C. Lee, G.~Chang, B.~Wang,
  N.~Alidoust, G.~Bian, M.~Neupane, C.~Zhang, S.~Jia, A.~Bansil, H.~Lin, and
  M.~Z. Hasan, \enquote{\protect{A Weyl Fermion semimetal with surface Fermi
  arcs in the transition metal monopnictide TaAs class},} Nature Communications
  \textbf{6}, 7373 (2015).

\bibitem{ref5}
Lv, B. Q. and Weng, H. M. and Fu, B. B. and Wang, X. P. and Miao, H. and Ma, J. and Richard, P. and Huang, X. C. and Zhao, L. X. and Chen, G. F. and Fang, Z. and Dai, X. and Qian, T. and Ding, H., \enquote{\protect{Experimental discovery of Weyl
  semimetal TaAs},} Physical Review X \textbf{5}, 031013 (2015).

\bibitem{ref6}
S.-Y. Xu, I.~Belopolski, N.~Alidoust, M.~Neupane, G.~Bian, C.~Zhang, R.~Sankar,
  G.~Chang, Z.~Yuan, C.-C. Lee, S.-M. Huang, H.~Zheng, J.~Ma, D.~S. Sanchez,
  B.~Wang, A.~Bansil, F.~Chou, P.~P. Shibayev, H.~Lin, S.~Jia, and M.~Z. Hasan,
  \enquote{\protect{Discovery of a Weyl fermion semimetal and topological Fermi
  arcs},} Science \textbf{349}, 613--617 (2015).

\bibitem{ref7}
W.~J. Chen, M.~Xiao, and C.~T. Chan, \enquote{\protect{Photonic crystals
  possessing multiple Weyl points and the experimental observation of robust
  surface states},} Nature Communications \textbf{7}, 13038 (2016).

\bibitem{ref8}
Q.~Lin, M.~Xiao, L.~Yuan, and S.~Fan, \enquote{\protect{Photonic Weyl point in
  a two-dimensional resonator lattice with a synthetic frequency dimension},}
  Nature Communications \textbf{7}, 13731 (2016).

\bibitem{ref9}
L.~Lu, L.~Fu, J.~D. Joannopoulos, and M.~Soljačić, \enquote{Weyl points and
  line nodes in gyroid photonic crystals,} Nature Photonics \textbf{7},
  294--299 (2013).

\bibitem{ref10}
L.~Lu, Z.~Wang, D.~Ye, L.~Ran, L.~Fu, J.~D. Joannopoulos, and M.~Soljačić,
  \enquote{\protect{Experimental observation of Weyl points},} Science
  \textbf{349}, 622--624 (2015).

\bibitem{gao2016photonic}
W.~Gao, B.~Yang, M.~Lawrence, F.~Fang, B.~B{\'e}ri, and S.~Zhang,
  \enquote{Photonic weyl degeneracies in magnetized plasma,} Nature
  communications \textbf{7} (2016).

\bibitem{goldman2013direct}
N.~Goldman, J.~Dalibard, A.~Dauphin, F.~Gerbier, M.~Lewenstein, P.~Zoller, and
  I.~B. Spielman, \enquote{Direct imaging of topological edge states in
  cold-atom systems,} Proceedings of the National Academy of Sciences
  \textbf{110}, 6736--6741 (2013).

\bibitem{celi2014synthetic}
A.~Celi, P.~Massignan, J.~Ruseckas, N.~Goldman, I.~B. Spielman,
  G.~Juzeli{\=u}nas, and M.~Lewenstein, \enquote{Synthetic gauge fields in
  synthetic dimensions,} Physical Review Letters \textbf{112}, 043001 (2014).

\bibitem{nalitov2015polariton}
A.~Nalitov, D.~Solnyshkov, and G.~Malpuech, \enquote{\protect{Polariton Z
  topological insulator},} Physical Review Letters \textbf{114}, 116401 (2015).

\bibitem{ref11}
S.~Ganeshan and S.~D. Sarma, \enquote{\protect{Constructing a Weyl semimetal by
  stacking one-dimensional topological phases},} Physical Review B \textbf{91},
  125438 (2015).

\bibitem{ref12}
Z.~Lan, N.~Goldman, A.~Bermudez, W.~Lu, and P.~Öhberg,
  \enquote{\protect{Dirac-Weyl fermions with arbitrary spin in two-dimensional
  optical superlattices},} Physical Review B \textbf{84}, 165115 (2011).

\bibitem{PhysRevX.7.041026}
A.~Weststr\"om and T.~Ojanen, \enquote{Designer curved-space geometry for
  relativistic fermions in weyl metamaterials,} Physical Review X \textbf{7},
  041026 (2017).

\bibitem{ref13}
M.~Saffman, T.~G. Walker, and K.~Mølmer, \enquote{\protect{Quantum information
  with Rydberg atoms},} Reviews of Modern Physics \textbf{82}, 2313 (2010).

\bibitem{gibbs2012optical}
H.~Gibbs, \emph{Optical bistability: controlling light with light} (Elsevier,
  2012).

\bibitem{ref15}
Y.~V. Kartashov, B.~A. Malomed, and L.~Torner, \enquote{Solitons in nonlinear
  lattices,} Reviews of Modern Physics \textbf{83}, 247 (2011).

\bibitem{lederer2008discrete}
F.~Lederer, G.~I. Stegeman, D.~N. Christodoulides, G.~Assanto, M.~Segev, and
  Y.~Silberberg, \enquote{Discrete solitons in optics,} Physics Reports
  \textbf{463}, 1--126 (2008).

\bibitem{dauxois2006physics}
T.~Dauxois and M.~Peyrard, \emph{Physics of solitons} (Cambridge University, 2006).

\bibitem{yang2010nonlinear}
J.~Yang, \emph{Nonlinear waves in integrable and nonintegrable systems} (SIAM,
  2010).

\bibitem{ref17}
Y.-R. Shen, \enquote{The principles of nonlinear optics,} New York,
  Wiley-Interscience \textbf{1}, 575 (1984).

\bibitem{ref18}
E.~J. Meier, F.~A. An, and B.~Gadway, \enquote{\protect{Observation of the
  topological soliton state in the Su-Schrieffer-Heeger model},} Nature
  Communications \textbf{7}, 13986 (2016).

\bibitem{ref19}
D.~Leykam and Y.~D. Chong, \enquote{Edge solitons in nonlinear-photonic
  topological insulators,} Physical Review Letters \textbf{117}, 143901 (2016).

\bibitem{kartashov2016modulational}
Y.~V. Kartashov and D.~V. Skryabin, \enquote{Modulational instability and
  solitary waves in polariton topological insulators,} Optica \textbf{3},
  1228--1236 (2016).

\bibitem{quiroga1997stable}
M.~Quiroga-Teixeiro and H.~Michinel, \enquote{Stable azimuthal stationary state
  in quintic nonlinear optical media,} JOSA B \textbf{14}, 2004--2009 (1997).

\bibitem{kevrekidis2005vector}
P.~Kevrekidis, H.~Susanto, R.~Carretero-Gonz{\'a}lez, B.~Malomed, and
  D.~Frantzeskakis, \enquote{Vector solitons with an embedded domain wall,}
  Physical Review E \textbf{72}, 066604 (2005).

\bibitem{carretero2008nonlinear}
R.~Carretero-Gonz{\'a}lez, D.~Frantzeskakis, and P.~Kevrekidis,
  \enquote{\protect{Nonlinear waves in Bose--Einstein condensates: physical
  relevance and mathematical techniques},} Nonlinearity \textbf{21}, R139
  (2008).

\bibitem{abdullaev2008localized}
F.~K. Abdullaev, A.~Gammal, M.~Salerno, and L.~Tomio,
  \enquote{\protect{Localized modes of binary mixtures of Bose-Einstein
  condensates in nonlinear optical lattices},} Physical Review A \textbf{77},
  023615 (2008).

\bibitem{reyna2014two}
A.~S. Reyna, K.~C. Jorge, and C.~B. de~Ara{\'u}jo, \enquote{Two-dimensional
  solitons in a quintic-septimal medium,} Physical Review A \textbf{90}, 063835
  (2014).

\bibitem{PhysRevLett.118.230403}
G.~Spagnolli, G.~Semeghini, L.~Masi, G.~Ferioli, A.~Trenkwalder, S.~Coop,
  M.~Landini, L.~Pezz\`e, G.~Modugno, M.~Inguscio, A.~Smerzi, and M.~Fattori,
  \enquote{Crossing over from attractive to repulsive interactions in a
  tunneling bosonic josephson junction,} Physical Review Letters \textbf{118},
  230403 (2017).

\bibitem{lin2009synthetic}
Y.-J. Lin, R.~L. Compton, K.~Jimenez-Garcia, J.~V. Porto, and I.~B. Spielman,
  \enquote{Synthetic magnetic fields for ultracold neutral atoms,} Nature
  \textbf{462}, 628--632 (2009).

\bibitem{dalibard2011colloquium}
J.~Dalibard, F.~Gerbier, G.~Juzeli{\=u}nas, and P.~{\"O}hberg,
  \enquote{Colloquium: Artificial gauge potentials for neutral atoms,} Reviews
  of Modern Physics \textbf{83}, 1523 (2011).

\bibitem{ref25}
J.~Cuevas-Maraver, P.~G. Kevrekidis, A.~Saxena, A.~Comech, and R.~Lan,
  \enquote{\protect{Stability of solitary waves and vortices in a 2D nonlinear
  Dirac Model},} Physical Review Letters \textbf{116}, 214101 (2016).

\bibitem{zou2016traveling}
P.~Zou, J.~Brand, X.-J. Liu, and H.~Hu, \enquote{Traveling majorana solitons in
  a low-dimensional spin-orbit-coupled fermi superfluid,} Physical Review
  Letters \textbf{117}, 225302 (2016).

\bibitem{ref20}
M.~S. Rudner and L.~S. Levitov, \enquote{\protect{Topological transition in a
  non-Hermitian quantum walk},} Physical Review Letters \textbf{102}, 065703
  (2009).

\bibitem{ref21}
D.~R. Hofstadter, \enquote{\protect{Energy levels and wave functions of Bloch
  electrons in rational and irrational magnetic fields},} Physical Review B
  \textbf{14}, 2239 (1976).

\bibitem{ref22}
T.~Dubcek, C.~J. Kennedy, L.~Lu, W.~Ketterle, M.~Soljacic, and H.~Buljan,
  \enquote{Weyl points in three-dimensional optical lattices: Synthetic
  magnetic monopoles in momentum space,} Physical Review Letters \textbf{114},
  225301 (2015).

\bibitem{RevModPhys.89.011004}
A.~Eckardt, \enquote{Colloquium: Atomic quantum gases in periodically driven
  optical lattices,} Reviews of Modern Physics \textbf{89}, 011004 (2017).

\bibitem{jaksch2003creation}
D.~Jaksch and P.~Zoller, \enquote{Creation of effective magnetic fields in
  optical lattices: the hofstadter butterfly for cold neutral atoms,} New
  Journal of Physics \textbf{5}, 56 (2003).

\bibitem{miyake2013realizing}
H.~Miyake, G.~A. Siviloglou, C.~J. Kennedy, W.~C. Burton, and W.~Ketterle,
  \enquote{Realizing the harper hamiltonian with laser-assisted tunneling in
  optical lattices,} Physical Review Letters \textbf{111}, 185302 (2013).

\bibitem{qi2011bound}
R.~Qi and H.~Zhai, \enquote{Bound states and scattering resonances induced by
  spatially modulated interactions,} Physical Review Letters \textbf{106},
  163201 (2011).

\bibitem{cappellini2014direct}
Cappellini, G. and Mancini, M. and Pagano, G. and Lombardi, P. and Livi, L. and Siciliani de Cumis, M. and Cancio, P. and Pizzocaro, M. and Calonico, D. and Levi, F. and Sias, C. and Catani, J. and Inguscio, M. and Fallani, L., \enquote{Direct
  observation of coherent interorbital spin-exchange dynamics,} Physical Review
  Letters \textbf{113}, 120402 (2014).

\bibitem{vortex}
Bleu, O and Malpuech, G and Solnyshkov, DD, \enquote{$\mathbb{Z}_2$ topological insulator analog for vortices in an interacting bosonic quantum fluid,} arXiv:1709.01830 (2017).



\bibitem{he2012berry}
Y.~He, J.~Moore, and C.~Varma, \enquote{Berry phase and anomalous hall effect
  in a three-orbital tight-binding hamiltonian,} Physical Review B \textbf{85},
  155106 (2012).

\bibitem{ref23}
K.~W. Kim, W.-R. Lee, Y.~B. Kim, and K.~Park, \enquote{\protect{Surface to bulk
  Fermi arcs via Weyl nodes as topological defects},} Nature Communications
  \textbf{7}, 13489 (2016).

\bibitem{ref24}
R.~Golizadeh-Mojarad and S.~Datta, \enquote{\protect{Nonequilibrium Green’s
  function based models for dephasing in quantum transport},} Physical Review B
  \textbf{75}, 081301 (2007).

\bibitem{huang2016experimental}
L.~Huang, Z.~Meng, P.~Wang, P.~Peng, S.-L. Zhang, L.~Chen, D.~Li, Q.~Zhou, and
  J.~Zhang, \enquote{Experimental realization of two-dimensional synthetic
  spin-orbit coupling in ultracold fermi gases,} Nature Physics \textbf{12},
  540--544 (2016).

\bibitem{meng2016experimental}
Z.~Meng, L.~Huang, P.~Peng, D.~Li, L.~Chen, Y.~Xu, C.~Zhang, P.~Wang, and
  J.~Zhang, \enquote{Experimental observation of a topological band gap opening
  in ultracold fermi gases with two-dimensional spin-orbit coupling,} Physical
  Review Letters \textbf{117}, 235304 (2016).

\bibitem{burkov2011weyl}
A.~Burkov and L.~Balents, \enquote{Weyl semimetal in a topological insulator
  multilayer,} Physical Review Letters \textbf{107}, 127205 (2011).

\bibitem{xiao2015synthetic}
M.~Xiao, W.-J. Chen, W.-Y. He, and C.~T. Chan, \enquote{Synthetic gauge flux
  and weyl points in acoustic systems,} Nature Physics \textbf{11}, 920--924
  (2015).

\bibitem{kim2015dirac}
Y.~Kim, B.~J. Wieder, C.~Kane, and A.~M. Rappe, \enquote{Dirac line nodes in
  inversion-symmetric crystals,} Physical Review Letters \textbf{115}, 036806
  (2015).

\bibitem{mullen2015line}
K.~Mullen, B.~Uchoa, and D.~T. Glatzhofer, \enquote{Line of dirac nodes in
  hyperhoneycomb lattices,} Physical Review Letters \textbf{115}, 026403
  (2015).

\bibitem{lim2017pseudospin}
L.-K. Lim and R.~Moessner, \enquote{Pseudospin vortex ring with a nodal line in
  three dimensions,} Physical Review Letters \textbf{118}, 016401 (2017).

\bibitem{ref26}
S.~Kumar, A.~M. Perego, and K.~Staliunas, \enquote{\protect{Linear and
  nonlinear bullets of the Bogoliubov-de Gennes excitations},} Physical Review
  Letters \textbf{118}, 044103 (2017).

\bibitem{ref27}
T.~R. Melvin, A.~R. Champneys, P.~G. Kevrekidis, and J.~Cuevas,
  \enquote{\protect{Radiationless traveling waves in saturable nonlinear
  Schrodinger lattices},} Physical Review Letters \textbf{97}, 124101 (2006).

\bibitem{li2008matter}
K.~Li, L.~Deng, E.~W. Hagley, M.~G. Payne, and M.~Zhan, \enquote{Matter-wave
  self-imaging by atomic center-of-mass motion induced interference,} Physical
  Review Letters \textbf{101}, 250401 (2008).

\bibitem{zhu2011strong}
J.~Zhu, G.~Dong, M.~N. Shneider, and W.~Zhang, \enquote{Strong local-field
  effect on the dynamics of a dilute atomic gas irradiated by two
  counterpropagating optical fields: Beyond standard optical lattices,}
  Physical Review Letters \textbf{106}, 210403 (2011).

\bibitem{dong2013polaritonic}
G.~Dong, J.~Zhu, W.~Zhang, and B.~A. Malomed, \enquote{Polaritonic solitons in
  a bose-einstein condensate trapped in a soft optical lattice,} Physical
  Review Letters \textbf{110}, 250401 (2013).

\bibitem{christodoulides1988discrete}
D.~Christodoulides and R.~Joseph, \enquote{Discrete self-focusing in nonlinear
  arrays of coupled waveguides,} Optics Letters \textbf{13}, 794--796 (1988).

\bibitem{hudock2004anisotropic}
J.~Hudock, N.~K. Efremidis, and D.~N. Christodoulides, \enquote{Anisotropic
  diffraction and elliptic discrete solitons in two-dimensional waveguide
  arrays,} Optics Letters \textbf{29}, 268--270 (2004).


\bibitem{ref28}
T.~I. Vanhala, T.~Siro, L.~Liang, M.~Troyer, A.~Harju, and P.~Törmä,
  \enquote{\protect{Topological phase transitions in the repulsively
  interacting Haldane-Hubbard model},} Physical Review Letters \textbf{116},
  225305 (2016).

\bibitem{yagasaki2005discrete}
K.~Yagasaki, A.~R. Champneys, and B.~A. Malomed, \enquote{Discrete embedded
  solitons,} Nonlinearity \textbf{18}, 2591 (2005).

\bibitem{meier2004experimental}
J.~Meier, G.~Stegeman, D.~Christodoulides, Y.~Silberberg, R.~Morandotti,
  H.~Yang, G.~Salamo, M.~Sorel, and J.~Aitchison, \enquote{Experimental
  observation of discrete modulational instability,} Physical Review Letters
  \textbf{92}, 163902 (2004).

\bibitem{ablowitz2011nonlinear}
M.~J. Ablowitz, \emph{Nonlinear dispersive waves: asymptotic analysis and
  solitons}, vol.~47 (Cambridge University, 2011).

\bibitem{dreisow2010classical}
F.~Dreisow, M.~Heinrich, R.~Keil, A.~T{\"u}nnermann, S.~Nolte, S.~Longhi, and
  A.~Szameit, \enquote{\protect{Classical simulation of relativistic
  Zitterbewegung in photonic lattices},} Physical Review Letters \textbf{105},
  143902 (2010).

\bibitem{keil2015optical}
R.~Keil, C.~Noh, A.~Rai, S.~St{\"u}tzer, S.~Nolte, D.~G. Angelakis, and
  A.~Szameit, \enquote{\protect{Optical simulation of charge conservation
  violation and Majorana dynamics},} Optica \textbf{2}, 454--459 (2015).

\bibitem{oreg2010helical}
Y.~Oreg, G.~Refael, and F.~von Oppen, \enquote{Helical liquids and majorana
  bound states in quantum wires,} Physical Review Letters \textbf{105}, 177002
  (2010).

\bibitem{das2012zero}
A.~Das, Y.~Ronen, Y.~Most, Y.~Oreg, M.~Heiblum, and H.~Shtrikman,
  \enquote{Zero-bias peaks and splitting in an al-inas nanowire topological
  superconductor as a signature of majorana fermions,} Nature Physics
  \textbf{8}, 887 (2012).

\end{thebibliography}
\end{document}